\documentclass[conference, 11pt, a4paper]{llncs}

\usepackage{graphicx}
\usepackage{subfigure}
\usepackage{amsmath, amssymb}
\usepackage{algorithm}
\usepackage{algpseudocode}

\begin{document}

\title{Improved Time Complexity of Bandwidth Approximation in Dense Graphs}

\author{Hao-Hsiang Hung\\
hhung2@emory.edu, hhung7@gatech.edu}
\institute{Emory University}
\date{}
\maketitle


\begin{abstract}

Given a graph $G=(V, E)$ and and a proper labeling $f$ from $V$ to 
$\{1, ..., n\}$, we define $B(f)$ as the maximum absolute difference 
between $f(u)$ and $f(v)$ where $(u,v)\in E$. 
The bandwidth of $G$ is the minimum $B(f)$ for all $f$.
Say $G$ is $\delta$-dense if its minimum degree is $\delta n$.
In this paper, we investigate the trade-off between the approximation ratio and the time complexity of the classical approach of Karpinski {et al}.\cite{Karpin97}, and present a faster randomized algorithm for approximating 
the bandwidth of $\delta$-dense graphs.
In particular, by removing the polylog factor of the time complexity required to enumerate all possible
placements for balls to bins, we reduce the time complexity from $O(n^6\cdot (\log n)^{O(1)})$ 
to $O(n^{4+o(1)})$.
In advance, we reformulate the perfect matching phase of the algorithm with a maximum flow problem of smaller size and reduce the time complexity to $O(n^2\log\log n)$.
We also extend the graph classes could be applied by the original approach: we show that the algorithm remains polynomial time as long as $\delta$ is $O({(\log\log n)}^2 / {\log n})$.

\end{abstract}

\section{Introduction}
The bandwidth problem has long been studied for its massive applications in layout design, linear equations solving, interconnection networks, constraint satisfaction problems, channel assignment, and bioinformatics \cite{Diaz02,Petit11,Lai99,Khen08}.
It was first investigated by Harper and Hales from the Jet Propulsion Laboratory in 1962.
In particular, they tried to find a scheme to minimize the maximum absolute error of the 6-bit picture codes on a hypercube.

We may define the bandwidth optimization problems as follows.
Given a graph $G=(V, E)$ and and a proper labeling $f$ from $V$ to $\{1, ..., n\}$, we define $B(f)$ as the maximum absolute difference between $f(u)$ and $f(v)$ where $(u,v)\in E$.
The \emph{Bandwidth} of $G$ (denoted as $B(G)$) is the minimum $B(f)$ for all $f$.

In complexity theory, the bandwidth problem is one of those notoriously hard problems -- it is NP-complete \cite{Pa76}.
Few graph classes have polynomial time algorithms for bandwidth \cite{Smithline95,Golovach11}.
In particular, even if the input graphs are restricted subclasses of trees (e.g., trees of maximum degree 3, or caterpillars with hair length at most 3), it remains NP-complete to compute the bandwidth \cite{Garey78,hara91}.
To determine the exact bandwidth of general graphs, researchers study exponential time algorithms.
The base of the exponents of algorithms has been improved from $O(10^n)$ to $O(4.83^n)$ \cite{DBLP:journals/tcs/CyganP10}.
On the other hand, if parameter $k$ is fixed, it takes polynomial time to verify if $k$ is the bandwidth of the graph \cite{Saxe80}.

Not only is it hard to compute the exact bandwidth of general graphs, in fact, to approximate the bandwidth even for caterpillars is APX-hard \cite{hara91,Feige09,DBLP:journals/jcss/DubeyFU11}.
For general graphs, Feige introduced an $O(\log^{3.5}n)$ approximation algorithms via the technique of volume respecting embeddings \cite{Feige00}, and Gupta improved it to $O(\log^{2.5}n)$ in trees and chordal graphs \cite{Gupta00}.
His original idea comes from the small distortion embeddings of Bourgain \cite{Bour85} and of Linial, London and Rabinovich \cite{Linial95}. 
Besides, Semidefinite relaxation has been investigated in \cite{Blum00} to provide approximation ratio of $O(\sqrt{n/{B(G)}}\log n)$.
For the connection between approximation ratio and fixed parameter tractability, it is proved that there is no $EPTAS$ for the bandwidth problem unless $FPT=W[k]$ (for all $k\geq 1$) \cite{karpeptas}.
D\'{i}az et al. \cite{Diaz01} analyze the relation between approximation ratio and good expansion properties in $G(n,p)$ models.

Among graph properties, the \emph{density} has attracted attention because of the success in approximating covering problems \cite{Karp96}, topological bandwidth \cite{Wirt98}, and a series of important NP-hard problems \cite{Karpin97,Arora95,Feder91} in dense graphs.
In 1997, Karpinski, Wirtgen, and Zelikovsky proposed a randomized algorithm to approximate bandwidth in $\delta$-dense graphs (please see the definition below) \cite{Karpin97}.
The idea includes several levels, combining techniques for problems about balls and bins, randomized dominating set, and matching \cite{Lovasz86,MVV87,Karp98}.
Their approach mainly relies on lemmas showing that for a given $\delta$-dense graph $G$, if we choose a random set $R\subset V$ with size $O(\log n)$, then with high probability $R$ is a dominating set, meaning every vertex is either in $R$, or has a neighbor in $R$.

According to the rough estimation \cite{Karpin97}, the time complexity of this 3-approximation algorithm is $O(|V|\cdot|E|\cdot PM(G)\cdot BIN(G^\prime))$, where $PM(G)$ is the time required for computing a perfect matching, and $BIN(G^\prime)$ is the time complexity of finding the optimal placement for "roots" (will be explained later) to the right boxes.
To find the best matching, they suggested the standard $s-t$ flow technique from \cite{Lovasz86} with time complexity $O(|V||E|)$, and also mentioned some space for improvement \cite{Karp98,MVV87}.
However, the time required for $BIN(G^\prime)$ is the dominating factor that makes improving  the perfect matching part less important, because they wanted to guarantee these "roots" to be put in the correct boxes, and it is not clear if there is an obvious way to avoid trying all possible box sizes, and eliminating all possible placements of the roots to find the optimal solution.

Suppose $r$ is the number of roots, then it takes $O(c^r)$ to compute the optimal layout by brute force (here we denote $n=|V|$, and $c$ is some constant).
The value of $r$ in \cite{Karpin97} is $O(\log n)$, therefore the time complexity of computing to optimal packing of roots into the boxes is $O(1/\delta)^{O(\log n)}=n^{O(\log 1/\delta)}$.
A dense graph has $\Omega(n^2)$ edges, therefore the total time complexity is $O(|V|\cdot|E|\cdot PM(G)\cdot BIN(G^\prime))=O((n\cdot n^2)^2\cdot (\log n)^{O(1)})=O((n^6)\cdot (\log n)^{O(1)})$.

In this paper we investigate Karpinski {et al.}'s approach \cite{Karpin97}, present an algorithm with improved time complexity, and extend the applicability of the algorithm to more general graph classes.
To improve the time complexity, we replace a polynomial factor with a polylog one by allowing small loss in the approximation ratio.
We target on the dominating factor which slows down the original scheme: the root selection phase, and we also slightly improve the perfect matching part by applying classical approaches \cite{MVV87,Karp98}.
In \cite{Karpin97}, the selection of the root set is important because its size determines if the exhaustive search takes exponential time or not.
Inspired from \cite{Vincent02}, in approximating the bandwidth of $\delta$-dense graphs, we studied the trade-off between the approximation ratio and the time complexity.
In particular, we focus on reducing size of the randomized selection procedure.
We show that by allowing a slight relaxation of the approximation ratio (with a constant factor), we could either obviously improve the time complexity, or extend the application to larger classes of graphs ($\delta$ does not have to be a constant; it can be $O({{(\log\log n)}^2/\log n})$.

\section{Preliminary}

\begin{definition}\textbf{Bandwidth}.
Given a graph $G=(V, E)$ and and a proper labeling $f$ from $V$ to 
$\{1, ..., n\}$, we define $B(f)$ as the maximum absolute difference 
between $f(u)$ and $f(v)$ where $(u,v)\in E$.
The Bandwidth of $G$ is the minimum $B(f)$ for all $f$, and we denote it as $B(G)$.
\end{definition}

Given an optimal labeling of a graph $G$ along a line with sorted vertices, we can assume any $v\in V$ has at most $2\cdot B(G)$ neighbors (otherwise there are not enough integers for labeling the neighbors of $v$).

Suppose the vertices are laid out according to $f$ and are partitioned into boxes of same size $B\geq B(f)$. 
Then any vertex in box $i$ has neighbors only in boxes $i-1$, $i$, or $i+1$.

\begin{definition}\textbf{Dense Graphs}.
We call a graph $G=(V, E)$ $\delta$-dense if its minimum degree is $\delta n$ for some constant $0<\delta<1$, where $n=|V|$.
\end{definition}
Note that if some vertex in $G$ has degree $D$, then any labeling $f$ has bandwidth $B(f)\geq D/2$.
In particular, any $\delta$-dense graph has $B(G)\geq\delta\cdot n/2$.
Conversely, any layout of the vertices into boxes of size $B$ so that all edges are restricted to adjacent boxes gives us a layout of $G$ with bandwidth $\leq 2\cdot B$.
\begin{corollary}
The optimal labeling of a $\delta$-dense graph has $O(1/\delta)$ boxes.
\end{corollary}

\begin{definition}\textbf{Distance Function}.
Given a graph $G=(V, E)$, the distance function $d:V\times V\rightarrow N$, where $N$ is the set of natural numbers, is defined by the number of edges in the shortest path between given two vertices.
Vertex $u$ is said to be $h$-hop to vertex $v$ iff $d(u,v)=h$.
\end{definition}
\begin{definition}\textbf{Distance Function between a Vertex and a Set}.
The distance function $d:V\times 2^V\rightarrow N$ is the number of edges in the shortest path between the given vertex and one vertex of the given set.
Vertex $u$ is said to be $h$-hop to set $S$ iff $d(u,S)=h$.
\end{definition}
\begin{definition}\textbf{Dominating Set}.
Given a graph $G=(V, E)$, $D_k\subseteq V$ is a $h$-dominating set iff
\newline
\begin{center}
$\forall v\in V, \exists u\in D_k$ such that $d(u,v)\leq h$,
\end{center}
that is, any node $v\in V$ is either in $D_h$ or at most $h$-hop to a node in $D_h$. 
The nodes in the dominating set are dominating nodes, and the others are non-dominating nodes.
If $u\in D_h$ and $d(u,v)\leq h$, then we call $u$ a dominator of $v$.
\end{definition}
Suppose $R\subseteq V$ is a dominating set and every $v\in R$ is has a neighbor $u\in R^\prime$ such that $R^\prime \subseteq V$.
Then $R^\prime$ is a 2-dominating set because apparently every vertex in $R$ is dominated by a vertex in $R^\prime$, and thus the distance between any vertex in $G$ to $R^\prime$ is at most 2.

We describe our approach in the following section.

\section{Our approach}\label{sec:app}
Our contribution in this paper has two different directions. We sketch them as follows.

First we relax the approximation ratio with a constant factor to significantly improve the time complexity of the algorithm \cite{Karpin97}.
To do this, we use a random 2-dominating set instead of a random dominating set.
The probabilistic analysis of the size of both sets is simple (and will be shown later) but important because they belong to different scales: the size of a random dominating set is $O(\log n)$, but the size of a random 2-dominating set is $O(\log\log n)$.
Call this random 2-dominating set $R^\prime$.
The algorithm starts by looping through all possible box sizes for bandwidth that we are guessing, and within the loop we enumerate placements mapping the set into the boxes, create the auxiliary graph according to the placement of $R$, to provide positions for vertices not in $R^\prime$.
Such an enumeration guarantees the optimal layout will be checked (so that the optimal bandwidth is recorded).

For the loop of the second level, we handle vertices not in $R^\prime$.
By applying perfect matching to $G^\prime$, we allocate them into the best possible boxes, allowing some constant approximation ratio.
We improve the complexity of the perfect matching algorithm from $O(|V||E|)$ to $O(\sqrt{|V|}|E|)$ via the classical algorithms \cite{MVV87,Karp98}.

We describe our approach in Algorithm \ref{alg:Framework}.
We need two lemmas for showing the correctness of the algorithm.
\begin{algorithm}[htb]
  \caption{The approximation algorithm for bandwidth in dense graphs}
  \label{alg:Framework}
  \begin{algorithmic}[1]
    \Require
      A graph $G$ which is $\delta$-dense
      \State Randomly select a subset $R^\prime\subset V$ of size $O(\frac{1}{\delta}\log\log n)$;
      \For{boxsize from $\delta n$ to $n/2$}
      \State Prepare a layout with $\lceil n/boxsize \rceil$ boxes;
      \For{Each possible placement of vertices in $R^\prime$ to the boxes}
	\State Build a bipartite auxiliary-graph $G^\prime$: 
      \For{each vertex $v\in V$}
      \State Construct $I_v$ in $G^\prime$;
      \State Connect $v$ to all possible places in $I_v$;
      \EndFor 
      \State Run a perfect matching algorithm:
      \If{$\exists$ a perfect matching $M$ in $G^\prime$}
      \State return it (as a layout);
	\Else
      \State continue;
      \EndIf
      \EndFor
      \EndFor
  \end{algorithmic}
\end{algorithm}
\begin{lemma}
Let $0<\alpha<1$, and $c$ a constant.
Given a $\delta$-dense graph $G$, we choose $k$ and $k^\prime$ to meet the following requirement.
Let $R$ be a randomly chosen set from $G$ of size
\begin{center}
\begin{equation}\label{equ:k}
k=\dfrac{\log(n/\alpha)}{\log(1/(1-\delta))}=O(\log n)
\end{equation}
\end{center}
, and $R^\prime$ be a randomly chosen set from $G$ of size
\begin{center}
\begin{equation}\label{equ:kprime}
k^\prime=\dfrac{\log(k\cdot c/\alpha)}{\log(1/(1-\delta))}=O(\log\log n)
\end{equation}
\end{center}
such that the expected number of vertices in $V$ not dominated by any vertex in $R$ is bounded by $\alpha$, 
and the expected number of vertices in $R$ not dominated by any vertex in $R^\prime$ is also bounded by $\alpha$.
Then $R^\prime$ is a 2-dominating set with probability at least ${(1-\alpha)}^2$.
\end{lemma}
\begin{proof}
Because $G$ is $\delta$-dense, the probability any particular vertex $v$ is dominated by a randomly chosen vertex is at least $\delta$.
Since $R$ is chosen and independently, suppose $k=|R|$, the probability that $v$ is not dominated by any vertex in $R$ is at most $(1-\delta)^k$.
At this step, the goal is to choose $k$ so that $(1-\delta)^k<\alpha/n$. 
\newline
\begin{center}
$(1-\delta)^k n\leq\alpha$
\end{center}
\begin{center}
$n/\alpha\leq (1/(1-\delta))^k$
\end{center}
\begin{center}
$\dfrac{\log(n/\alpha)}{\log(1/(1-\delta))}\leq k$
\end{center}

One can easily check that by our choice of $k$, this is at most $\alpha$.
So by Markov's inequality $R$ is a dominating set with probability at least $1-\alpha$.

Next we prove that $R^\prime$ dominates $R$ by similar arguments.
That means $R^\prime$ is a 2-dominating set.
The goal is to choose $k^\prime$ so that $(1-\delta)^{k^\prime}<\alpha/k$.
\begin{center}
$(1-\delta)^{k^\prime}c\cdot\log n\leq\alpha$
\end{center}
\begin{center}
$(\log n)\cdot c/\alpha\leq (1/(1-\delta))^{k^\prime}$
\end{center}
\begin{center}
$\dfrac{\log((\log n)\cdot c/\alpha)}{\log(1/(1-\delta))}\leq k^{\prime}$
\end{center}
Since $R$ is a dominating set of $G$ with probability at least $(1-\alpha)$ and any vertex of $R$ is dominated by $R^\prime$ with probability at least $(1-\alpha)$, $R^\prime$ is a 2-hop dominating set with probability at least ${(1-\alpha)}^2$.

Note we use $R$ for the convenience of explanation.
In the algorithm we do not need $R$ but $R^\prime$.
\qed
\end{proof}


\subsection{Auxiliary Graph}\label{aux}
We need an \emph{auxiliary graph} from $G$ for labeling.
Basically it is a bipartite graph $G^\prime=(X\bigcup Y, E)$ such that $X=V$ and $Y$ is a collection of possible positions to place vertices.

Call $R^\prime$ the "roots" of the graph. 
First we place these roots into the boxes (of given size).
Since $G$ is $\delta$-dense, at most $O(1/\delta)$ boxes are needed.
Besides, there is no obvious way to cancel some impossible placements, so we try out all of them.

Secondly we build $G^\prime$ according to the placement of $R^\prime$.
For each $v\in V$, let $r_v$ be some vertex in $R^\prime$ at distance 2 from $v$. 
%
By Breadth-first Search with centers from $R^\prime$, we classify the vertices of $G^\prime$ into layers:
Let $L_1=\{v\in V-R^\prime | d(v, R^\prime)=1\}$, and let $L_2=\{v\in V-R^\prime-L_1 | d(v, L_1)=1\}$. 
For every $v\in L_1$ we record all $u\in R^\prime$ such that $d(v,u)=1$.
For every $w\in L_2$ we record all $v\in L_1$ such that $d(w,v)=1$.
Here we define $L_2$ based on $L_1$ for algorithmic consideration.

Denote $B(v)$ the index of the box in $G^\prime$ where $v$ is placed.
Actually, we can put $v$ in at most five different (and consecutive) boxes because $v$ and any $u\in\{r_v\}$ can be at most two boxes away, namely $|B(v)-B(u)|\leq 2$.
Call these boxes (in $G^\prime$) $I_v$.
That is, $I_v=\bigcap_{u\in R^\prime, d(u,v)\leq 2}\{B(u)-2,\cdots,B(u)+2\}$.
If $I_v$ is empty then no solution exists.
For each box $i\in I_v$, we connect $v$ to all the vertices in $i$ ($v$ can be placed in one of these positions). 

\subsection{Perfect Matching}
Given $G^\prime=(V^\prime, E^\prime)$, the next phase is a perfect matching algorithm.
If a perfect matching $M\subseteq E^\prime$ exists, we return the layout as a solution.
Otherwise we consider the next placement of vertices in $R^\prime$ (and reconstruct $G^\prime$).
If all the placements with current boxsize are examined and failed, we continue checking a larger size.
By the suggestion in \cite{Karpin97}, there are several candidates \cite{MVV87,Karp98} for improving the time complexity to $O(\sqrt{|V|}|E|)$.

\begin{lemma}
The approximation ratio of algorithm 1 is at most 10.
\end{lemma}
\begin{proof}
Consider two vertices $u$ and $v$ in the optimal layout ($(u,v)\in E$, and $B(u)=B(v)$ or $B(v)=B(u)+1$):
if both $u$ and $v$ are in $R^\prime$ then the layout is optimized by algorithm 1.

If one vertex dominates the other (without loss of generality we assume $u$ dominates $v$), then in worst case $v$ might be assigned $B(u)-1$ or $B(u)+1$, so the approximation ratio is 2.

If $u$ and $v$ do not have dominating relation, and $u$'s index is less than $v$'s in the labeling (without loss of generality), then in the worst case $r_u$ can be in box $B(u)-2$, and the matching algorithm assigns $u$ to box $B(u)-4$ (and $v$ to box $B(v)+4$).
Therefore $u$ and $v$ could be at most 10 boxes away from each other.\qed
\end{proof}
\begin{lemma}
Under careful analysis, the approximation ratio of algorithm 1 is at most 6.
\end{lemma}
\begin{proof}
Suppose $u$ and $v$ are two vertices with roots $r_u$ and $r_v$ and assume $(u,v)\in E$.
In $G^\prime$, we observe that $u$ can be at most three hops away from $r_v$ (and $d(v, r_u)\leq 3$, too).
Namely, $r_v$ dominates $u$ if we extend the domination relation to 3-hops.
We add these constraints when we construct $G^\prime$ to improve the approximation ratio.
\begin{figure}[ht]
\centering
\subfigure[Case 1: 6 approximation]{
\includegraphics[scale=0.13]{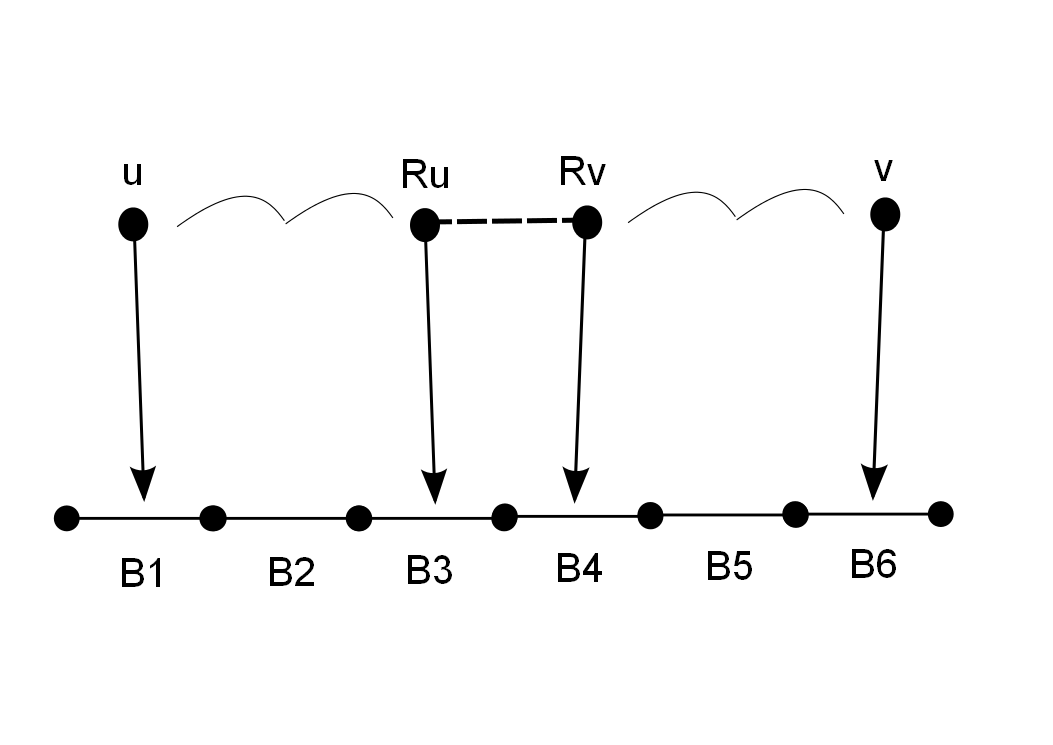} 
\label{fig:6app}
}
\subfigure[Case 2: 5 approximation]{
\includegraphics[scale=0.13]{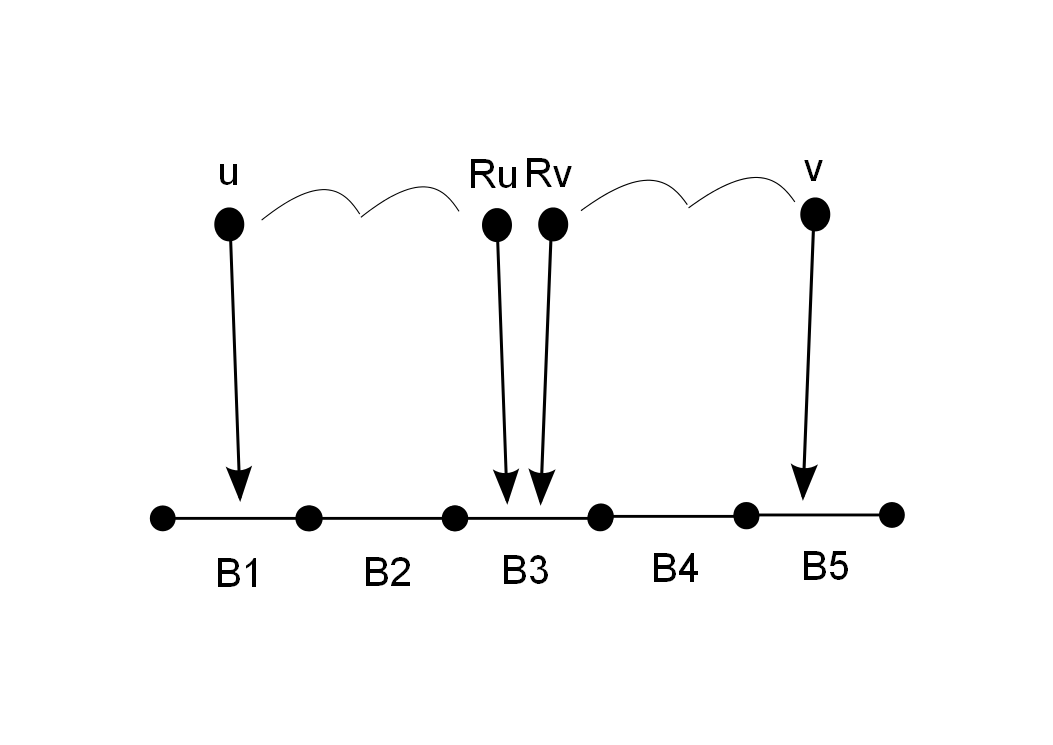} 
\label{fig:5app}
}
\label{worstapp}
\caption{The worst approximation ratio of Algorithm 1: (a) the case when $\gamma=1$, (b) the case when both roots are in the same box. The arcs show the placement and the dotted edges are real edges in 
$G$.}
\end{figure}

We discuss the approximation ratio according to the hop-distance $d(r_u,r_v)$.
Let $d(r_u, r_v)=\gamma$.
In the following cases, we assume $r_u$ is always in box $i$, and $r_v$ in box $i+\gamma$ ($\gamma\geq 0$ without loss of generality).
\begin{itemize}
\item[Case 1.] Suppose $1\leq\gamma\leq 5$.
Since $d(u, r_v)\leq 3$ and $d(v, r_u)\leq 3$, it is not hard to check that the leftmost possible position of $u$ could be in box $i+k-3$ and the rightmost possible position of $v$ could be in box $i+3$.
Therefore the approximation ratio is at most $7-\gamma$.
\item[Case 2.] Suppose $\gamma = 0$. The leftmost possible position of $u$ could be in box $i-2$, and the rightmost possible position of $v$ could be in box $i+2$. Therefore the approximation ratio is at most $5$.
\end{itemize}
\qed
\end{proof}
\begin{lemma}
The time complexity of algorithm 1 is $O(n^{5.5+o(1)})$. 
\end{lemma}
\begin{proof}
Selecting the randomized 2-dominating set $R^\prime$ takes $O(\log\log n)$, which is isolated from the rest loop procedure.

The outermost loop tests the range of the box size in $O(n)$.
Placing $R^\prime$ into boxes takes $O(\frac{1}{\delta})^{O(\frac{1}{\delta}\log\log n)}$ configurations because only $O(1/\delta)$ boxes are needed.
Building the auxiliary graph takes (including the construction of $I_v$) requires $O(n^2)$.
The perfect matching algorithm runs in $O(\sqrt{|V|}|E|)=O(n^{2.5})$. 
Multiply the above factors together we get $O(n^{5.5+o(1)})$.
\qed
\end{proof}
\begin{theorem}
Given a $\delta$-dense graph $G$, there is a $6$-approximation algorithm for the bandwidth problem of $G$ in polynomial time.
\end{theorem}
\begin{proof}
By Lemma 1, 3, and 4.\qed
\end{proof}

We could improve the time complexity of Algorithm 1 by the following analysis.

\begin{lemma}
The time complexity of algorithm 1 is $O(n^{4+o(1)})$. 
\end{lemma}
\begin{proof}
Observe that the auxiliary graph $G^\prime$ has two sides $X=V$ and $Y=B$, where $B$ is the collection of boxes with $|B|=O(1/\delta)$, so the number of vertices of $G^\prime$ is $O(n)$.
Each $x\in X$ connects to at most five different $y\in Y$, therefore the number of edges in $G^\prime$ is $O(n)$.
The best known time complexity of perfect matching takes $O(\sqrt{|V|}|E|)=O(n\sqrt{n})$ in this setting, so we can improve it to $O(n^{4+o(1)})$.\qed
\end{proof}

\section{Further Improvement}
In this section, we further investigate the matching procedure and provide a faster algorithm for the bandwidth problem, by re-formulating it into a flow problem of smaller size.
The ideas comes from a natural connection between bipartite matching and maximum flow problem.
We sketch the major approach as follows: we build a smaller bipartite graph to improve the time complexity of the matching: by attaching the source and sink nodes, we reduce it into a maximum flow problem.
We can solve it by any algorithm for the maximum flow problem and decide to continue testing or not by the results.
If the maximum flow exists (here we mean the units of flow out of the source equal to $|V|$), convert the solution back to a layout for the perfect matching; otherwise we continue testing (by using different configurations or enlarging boxsizes).
We describe it in Algorithm \ref{alg:better}.

For testing the boxsize, if the approximation ratio is allowed, then we do not have to find the exact size.
We can consider using binary search instead to reduce the time complexity of the outermost loop of the algorithm to $O(\log n)$.

\subsection{Construct Interval}
We first notice that the construction of the interval from Algorithm \ref{alg:Framework} can be removed out of the for loop: it can be maintained in a table and be updated more efficiently later.
The table can be created following the search procedure in the auxiliary graph, but to create a real auxiliary graph is not necessary. 
The rows list the vertices of $G$; the columns list vertices of $R^\prime$.
If some $v\in V$ is dominated by some $u\in R^\prime$, we record the start and end indices of boxes.
Updating the table for a different configuration can be done in polynomial time.

\begin{algorithm}[htb]
  \caption{Faster approximation algorithm for bandwidth in dense graphs}
  \label{alg:better}
  \begin{algorithmic}[1]
    \Require
      A graph $G$ which is $\delta$-dense
      \State Randomly select a subset $R^\prime\subset V$ of size $O(\frac{1}{\delta}\log\log n)$;
	 \State BFS starting from vertices in $R^\prime$;
      \For{boxsize from $\delta n$ to $n/2$}
      \State Prepare a layout with $\lceil n/boxsize \rceil$ boxes;
      \For{Each possible placement of vertices in $R^\prime$ to the boxes}
      \For{each vertex $v\in V$}
      \State Construct/Update $I_v$ (start and end indices specified);
      \EndFor
      \State Create the flow instance, counting $B_{ij}$ according to the intervals of $I_v$;
      \State Solve the flow problem by any maximum flow algorithm;
      \If{the flow value is equal to $|V|$}
      \State Convert the solution back to the perfect matching;
      \State Return it as a layout;
      \Else
      \State Otherwise continue (no perfect matching exists in such boxsize);
	\EndIf 
      \EndFor
      \EndFor
  \end{algorithmic}
\end{algorithm}
\subsection{Maximum Flow}


Our next step is to build a new bipartite graph by the collection of the intervals of vertices.
Denote $I_v$ be the collection of boxes that $v$ is allowed to occupy.
In other words, $I_v$ are boxes of an interval $\subseteq \{1,...,b\}$, where $b$ is the number of boxes.
Let $B_k$ be the box of index $k$. 
From Algorithm 1 we know how to build $I_v$ by intersecting the intervals representing the coverage all possible roots of $v$.
Knowing where a given root is placed, here by coverage we mean all the boxes containing the vertices dominated by that root within two hops.
In fact, we can demonstrate $I_v=\{i,\cdots,j\}$ by an index pair of starting and ending boxes, say $B_{ij}$.
The special case is when $v\in R^\prime$: now $i=j$, meaning it can only be placed in that specific box. 
There are at most $\binom{b}{2}$ distinct (and consecutive) intervals for representing all possible ranges (indeed 5b by Section \ref{aux}).

Define $c_{ij}=|\{v:I_v=\{i,\cdots,j\}\}|$, the number of vertices using interval $B_{ij}$.
To begin with, we count $c_{ij}$ by checking the index pairs of starting and ending boxes, with time complexity $O(b^2 n)$.
Then we build a bipartite graph $H=\{L\bigcup R, E\}$, where $L=\{B_{ij}\}$, and $R$ represent all the $b$ boxes.
For every $B_{ij}$ we build directed edges to box $i$ and to box $j$, and all the boxes in between.
We set infinite capacity to all of them.
A source node $s$ is added with directed edges to all $B_{ij}\in L$ with capacity $c_{ij}$ respectively.
Additionally, a sink node $t$ is added with directed edges from all $r\in R$, with capacity given by the boxsize of the outer loop.

We then solve the flow problem with any maximum flow algorithm, with no concern about the time complexity. 
This is because in the new setting the number of nodes and the number of edges are independent of $n$, the number of vertices of $H$.
Given a saturation flow $g$, we check if the total amount of the flow out from the source node is equal to the number of vertices in $G^\prime$.
If yes, then we convert $s$ back into an optimal solution of the original bipartite matching problem in Algorithm 1, for a layout of $G^\prime$ with optimal bandwidth.
Otherwise we continue testing by changing the placement of $R^\prime$, or enlarging the boxsize.

We describe the conversion procedure as follows.
For each $B_{ij}$, define $f_{ij}$ the value of the flow in and out of it (they should be equal by the conservation condition, required by the flow problem).
Then arbitrarily pick $f_{ij}$ vertices from $V$ (with $I_v=B_{ij}$) and place them to the boxes.

Let $\bar{g}$ be the converted solution.
\begin{lemma}
$\bar{g}$ is an optimal solution of the bipartite matching phase in Algorithm 1.
\end{lemma}
\begin{proof}
We prove that there is a saturation flow $g$ in Algorithm \ref{alg:better} with value $n$ if and only if there is a perfect matching $M$ of $G^\prime$ in Algorithm \ref{alg:Framework}.

Denote $f(u,v)$ as the total amount of flow from $u$ to $v$ along edge $(u,v)$.
We can represent the value of $g$ as $\sum_{i,j} f(s, B_{ij})$ (the total number of vertices selected between all possible intervals), or $\sum_{k} f(B_k, t)$ (the total number of vertices (belong to intervals) placed into boxes).
Note the capacities of the edges are either integers or infinite, so if a saturation flow exists then it is integral.
If there is a flow in edge $(B_{ij}, B_k)$ with unit $e$, then in Algorithm \ref{alg:better} we select $e$ vertices from the table with interval starting from $i$ to $j$, which means we place these $e$ vertices into spots in box $k$ until they are full.

Conversely, given a perfect matching $M$ of $G^\prime$ such that $m={v, B_k}$ for each $m\in M$, we can specify one unit of flow in $(s, B_{ij})$ where $B_v=B_{ij}$, and also reserve a unit of flow in $(B_{ij}, B_k)$ and in $(B_k, t)$. Since all the edges out of $s$ and into $t$ are saturated, it is a saturation flow. 
\qed
\end{proof}
\begin{lemma}
The time complexity of Algorithm \ref{alg:better} is $O(n^2\log\log n)$.
\end{lemma}
\begin{proof}
(Remark: since $G$ is dense, this is near linear.)
Before we eliminate all the possible placement of roots into boxes, the loop of boxsize has $O(n)$ configurations, and the worst case performance of the breadth first search is $O(|V|+|E|)=O(n^2)$, and $R^\prime=O(\log\log n)$, so the time complexity is $O(n^2\log\log n)$. 
The time complexity of creating the table is $O(n\log\log n)$ because it takes $O(1)$ to calculate the entry of the table from the search result.

There are $(\log n)^{O(\frac{1}{\delta}\log\frac{1}{\delta})}$ different configurations for placing roots into boxes.
We have to update $O(n\log\log n)$ entries of the table.
The creating of the smaller maximum flow instance needs to pair up $O(b^2n)$ cases, and solving it takes constant time (either for correct or incorrect output).
Converting a solution back to the bipartite matching and the layout can be done by checking the table with time complexity $O(n\log\log n)$.
Since all these four phases are independent, in this part the time complexity is $O(\log n\cdot n\log\log n\cdot(\log n)^{O(\frac{1}{\delta}\log\frac{1}{\delta})})$.

Actually, the time complexity of Algorithm \ref{alg:better} is dominated by the creation of the table.\qed 
\end{proof}
\begin{theorem}
Algorithm \ref{alg:better} is a 6-approximation algorithm for bandwidth problem in $\delta$-dense graphs with time complexity $O(n^2\log\log n)$.
\end{theorem}
\begin{proof}
By Lemma 6, 7.
\end{proof}
\subsection{Extend to Larger Graph Classes}
In this subsection we discuss the graph classes where Algorithm \ref{alg:Framework} could apply.
By Definition 2, $\delta$ is a constant.
With careful calculation, $\delta$ could be extended to $O(\frac{{(\log\log n)}^2}{\log n})$, and here the trade-off is we sacrifice some improvement of the time complexity.
We analyze such a case as follows.

If we analyze the equations (\ref{equ:k}) and (\ref{equ:kprime}) in Section \ref{sec:app} more carefully by assuming $\alpha\sim1$, we get $k=O(\frac{1}{\delta}\log\frac{1}{\delta})$ and $k^\prime=O(\frac{1}{\delta}[\log\frac{1}{\delta}+\log\log n])$.
Set $\delta=O(\frac{{(\log\log n)}^2}{\log n})$, we still get $k^\prime=O(\log\log n)$.
Since such $\delta$ depends on $n$, the dense graph classes are larger than the original assumption.
Therefore, we could either improve the time complexity of the placement of roots, or extend the results to larger dense graph classes.
\section{Conclusion and Open Problems}

We have considered $h$-hop dominating set with $h\geq 3$ for improving the performance of the algorithms.
The result is not very optimistic: $h\geq 3$ does not help much; $h=2$ is enough.
This is because when the scale of the dominating set becomes smaller (so is base of the time complexity of the brute force bin packing), $\delta$ will take over and dominate the overhead.
It should be more interesting to investigate other possible techniques to improve the time complexity of the algorithm, or to improve the approximation ratio and still to preserve the efficiency of our approach.

\section*{Acknowledgement}
I want to thank my patient advisor Professor Michelangelo Grigni for sharing his imaginative trip in math.
\bibliographystyle{abbrv}
\bibliography{band}

\end{document}